\documentclass[11pt]{article}
\usepackage{amssymb}
\usepackage{amsmath}
\usepackage{amstext}
\usepackage{graphicx,epsfig}
\usepackage{epsfig}
\usepackage{verbatim} 
\usepackage{fancybox}
\usepackage{color}
\usepackage{ulem}
\usepackage{enumitem}
\usepackage{subfigure}
\usepackage{bbm}
\usepackage{parskip}
\usepackage[numbers,sort&compress]{natbib}

\usepackage{tikz}
\usetikzlibrary{calc,decorations.markings}

\newcommand{\Comment}[1]{{}}
\definecolor{darkblue}{rgb}{0.15,0.35,0.55}
\definecolor{reddish}{rgb}{0.65, 0.2, 0.2}
\usepackage[linktocpage=true]{hyperref}
\hypersetup{
colorlinks=true,
citecolor=darkblue,
linkcolor=reddish,
urlcolor=darkblue,
pdfauthor={},
pdftitle={},
pdfsubject={}
}

\newcommand{\be}{\begin{equation}}
\newcommand{\ee}{\end{equation}}
\newcommand{\bea}{\begin{eqnarray}}
\newcommand{\eea}{\end{eqnarray}}
\newcommand{\beas}{\begin{eqnarray*}}
\newcommand{\eeas}{\end{eqnarray*}}

\def\({\left(}
\def\){\right)}

\def\gsim{ \lower .75ex \hbox{$\sim$} \llap{\raise .27ex \hbox{$>$}} }
\def\lsim{ \lower .75ex \hbox{$\sim$} \llap{\raise .27ex \hbox{$<$}} }

\def\xyma{\xymatrix@M.7em}
\def\xymas{\xymatrix@M.1em}
\newcommand{\ba}{\begin{eqnarray}}
\newcommand{\ea}{\end{eqnarray}}

\title{}
\author{}

\numberwithin{equation}{section}
\begin{document}
\def\thefootnote{\fnsymbol{footnote}}
\setcounter{footnote}{0}

\begin{center}
\LARGE{\textbf{Finite Temperature Description of an Interacting Bose Gas}} \\[0.5cm]

\vspace{.2cm}

\end{center}

\thispagestyle{empty}
\centerline{{\Large Anushrut Sharma,${}^{\rm a,}$\footnote{\href{mailto:anushrut@sas.upenn.edu}{\texttt{anushrut@sas.upenn.edu}}} Guram Kartvelishvili,${}^{\rm a,}$\footnote{\href{mailto:guramk@sas.upenn.edu}{\texttt{guramk@sas.upenn.edu}}} and Justin Khoury${}^{\rm a,}$\footnote{\href{mailto:jkhoury@sas.upenn.edu}{\texttt{jkhoury@sas.upenn.edu}}}}}
\vspace{.5cm}

\centerline{\it~$^{\rm a}$Center for Particle Cosmology, Department of Physics and Astronomy,}
\centerline{\it University of Pennsylvania, Philadelphia, PA 19104, USA}

\vspace{.6cm}

\hrule \vspace{0.2cm}
\centerline{\small{\bf Abstract}}

\vspace{-0.2cm}
{\small\noindent 
We derive the equation of state of a Bose gas with contact interactions using relativistic quantum field theory. The calculation accounts for both thermal and quantum corrections up to 1-loop order. We work in the Hartree-Fock-Bogoliubov approximation and follow Yukalov's prescription of introducing two chemical potentials, one for the condensed phase and another one for the excited phase, to circumvent the well-known Hohenberg-Martin dilemma. As a check on the formalism, we take the non-relativistic limit and reproduce the known results. Finally, we translate our results to the hydrodynamical, two-fluid model for finite-temperature superfluids. Our results are relevant for the phenomenology of Bose-Einstein Condensate and superfluid dark matter candidates, as well as the color-flavor locking phase of quark matter in neutron stars.}

\vspace{0.3cm}
\noindent
\hrule
\vspace{1.3cm}
\def\thefootnote{\arabic{footnote}}
\setcounter{footnote}{0}

For over two decades, Bose-Einstein Condensates (BEC) have been considered as a candidate for Dark Matter~\cite{Sin:1992bg, Ji:1994xh, Goodman:2000tg, Peebles:2000yy, Arbey:2003sj, Boehmer:2007um, Harko:2011xw, Harko:2014vya, Slepian:2011ev, Guth:2014hsa, 2011PhRvD..84d3531C}. In particular, there has been considerable interest in the theory of dark matter superfluidity~\cite{Berezhiani:2015bqa, Berezhiani:2015pia,Berezhiani:2017tth,Berezhiani:2020umi,Berezhiani:2021rjs}, which allows dark matter to realize various empirical galactic scaling relations, {\it e.g.},~\cite{Milgrom:1983ca,McGaugh:2011ac,McGaugh:2016leg,Lelli:2016cui,Salucci:2018hqu}. In the initial studies, the dark matter density profile was computed using a simplified equation of state valid at zero temperature. Subsequently, two of us computed the non-relativistic, finite-temperature equation of state for dark matter superfluids with 2- and 3-body contact interactions, using a self-consistent mean-field approximation~\cite{Sharma:2018ydn}. This allowed us to derive the finite-temperature density profile for dark matter superfluids, under the assumption of spherical symmetry and uniform temperature.

The goal of this paper is to revisit this calculation using a relativistic quantum field theory (QFT) framework. As usual, the upshot of an effective field theory approach is that it offers a systematic treatment of the relevant degrees of freedom and the symmetries governing their dynamics.  Although a complete relativistic calculation of a BEC/superfluid might not be of immediate relevance to dark matter, it is nevertheless important for various other phenomena, such as the color-flavor locking superfluid phase of quark matter~\cite{Alford:1998mk} conjectured to reside in the core of neutron stars. The associated observable phenomena, like pulsar glitches~\cite{1975Natur.256...25A}, would be sensitive to the relativistic corrections. 

To properly account for the depletion of a BEC with increasing temperature, one must perform a self-consistent calculation. In QFT, this requires working with the Cornwall-Tomboulis-Jackiw (CJT) or the two particle-irreducible (2PI) framework~\cite{Cornwall:1974vz}. In this approach, the effective action is computed in terms of the background field as well as the dressed propagator. The CJT formalism has also been extensively applied to thermal field theory~\cite{NORTON1975106, Amelino-Camelia:1992qfe, Millington:2012pf}. The 2PI effective action is expressed in terms of an infinite set of diagrams having partially resummed propagators. Thus, for practical purposes, one must truncate the 2PI effective action at finite order in the loop diagram expansion. This truncation, however, gives rise to residual violations of global symmetries, which prevent the Goldstone boson of the spontaneously broken phase from being gapless. The Euler-Lagrange equations of motion of the truncated effective action are not consistent with the Ward identities. On the other hand, the complete theory must be invariant under the global~$U(1)$ symmetry, and as such must give rise to a massless Goldstone mode~\cite{Goldstone:1962es}. The inability of the CJT effective action to simultaneously satisfy the mean-field equation and display a gapless mode is similar to the Hohenberg-Martin dilemma~\cite{HOHENBERG1965291} and has been noted in literature~\cite{Amelino-Camelia:1997xip, Petropoulos:1998gt, Lenaghan:1999si}.    

Various working solutions have been proposed to this problem depending on the particular application. It has been found that in~$O(2N)$ theories, a Goldstone mode can be recovered in the large-$N$ limit~\cite{Petropoulos:1998gt, Lenaghan:1999si, PhysRevD.75.065011}. Other solutions add a phenomenological or an {\it ad hoc} term to satisfy both constraints~\cite{PhysRevD.71.105016, PhysRevD.72.036008, Baacke:2002pi, Alford:2013koa}. Another way to obtain a massless Goldstone boson is to define a constrained version of the CJT formalism with the so-called external propagators~\cite{vanHees:2002bv, Nemoto:1999qf} by giving up on a second-order phase transition~\cite{Marko:2015gpa}. Lastly, the symmetry-improved CJT formalism introduces new constraints on the loop-truncated CJT effective action to enforce the Ward identities~\cite{Pilaftsis:2013xna}. While offering different insights into the nature of the problem, each of these approaches either carries a pathology, introduces {\it ad hoc} terms, or enforces constraints by hand.

For concreteness, in this paper, we follow the framework proposed by Yukalov~\cite{YUKALOV2008461}. This approach relies on introducing two different chemical potentials --- one for the condensed phase, 
and a second one for the normal phase. The two chemical potentials are distinct below the critical temperature, and allow us to simultaneously satisfy the self-consistency condition for the mean-field while having 
a gapless Goldstone mode. The two chemical potentials become equal at the critical temperature. Above the critical temperature, there is of course a single chemical potential,
associated with the conserved particle number. Like the other aforementioned approaches to the Hohenberg-Martin dilemma, the framework of~\cite{YUKALOV2008461} introduces an additional variable to account for the different constraints. One might argue, however, that the usage of two chemical potentials can be physically motivated, and can be naturally extended to a QFT exhibiting spontaneous symmetry breaking.

The paper is organized as follows. We first set the stage for the calculation by reviewing the imaginary time formalism and the Hartree-Fock-Bogoliubov (HFB) approximation~\cite{Bender:2003jk} in Sec.~\ref{section one-loop effective action}. We then introduce Yukalov's framework in Sec.~\ref{2 chemical potentials justification}, justifying the use of two chemical potentials. We will then work out the Matsubara summation in Sec.~\ref{Matsubara Summation} to compute the equation of state in Sec.~\ref{Relativistic Equation of State}, before computing the necessary correlators and renormalizing our theory in Sec.~\ref{Calculating the Expectation Values}. As a check on our results, we work out the non-relativistic limit of our theory in Sec.~\ref{Non-Relativistic Limit} and compare it to the results of~\cite{Sharma:2018ydn}. As another application, in Sec.~\ref{Hydrodynamics of a Superfluid}, we translate our results to the hydrodynamical two-fluid model of superfluidity~\cite{PhysRev.60.356,1938Natur.141..913T,1938Natur.141..643L}, in which the system is treated as a mixture of the superfluid component and a normal fluid, made up of quasi-particles.


\section{One-loop effective action}
\label{section one-loop effective action}

Similar to the grand partition function, the functional integral for a \text{complex scalar field}~$\Phi$, in the imaginary time formalism, is written as~\cite{Kapusta:2006pm} 
\be
Z= \int \mathcal{D}\Phi\mathcal{D}\Pi\ \exp\left[\int_{0}^{\beta} \text{d}\tau\int \text{d}^3x\Big(\rm{i}\Pi^\dagger \partial_{\tau}\Phi + \rm{i} \Pi\partial_{\tau}\Phi^\dagger -\mathcal{H} + \mu \it{N}\Big)\right]\,,
\label{Z one mu}
\ee
where~$\Pi$ is the conjugate momentum, and the field~$\Phi$ is periodic in imaginary time,~$\Phi(\vec{x},0) = \Phi(\vec{x},\beta)$, with~$\beta = 1/T$.\footnote{We work in natural units where Boltzmann's constant~$k_{\rm B}$ is set to unity, as well as~$\hbar = c = 1$.} The free energy density is given by
\be
F = \frac{1}{ V \beta}\ln Z\,,
\label{F def}
\ee
where~$V$ is the volume. For a complex scalar field, the Hamiltonian density~$\mathcal{H}$ is 
\be
\mathcal{H} =  \Pi^{\dagger}\Pi  + \vec{\nabla}\Phi^{\dagger}\cdot\vec{\nabla}\Phi + U(\Phi^\dagger, \Phi)\,, 
\ee
where the potential~$U(\Phi^\dagger, \Phi)$ is assumed to arise from self-interactions rather than coupling to an external field. Furthermore, we assume that the predominant interactions are particle-conserving contact interactions, in which case~$U(\Phi^\dagger, \Phi) = U(|\Phi|^2)$. Thus the potential~$U(\Phi^\dagger, \Phi)$ has a~$U(1)$ symmetry. The outline of the subsequent calculation will be valid for any such~$U(|\Phi|^2)$~\cite{Benson:1991nj}. For concreteness, however, we specialize to the simple potential
\be
U = m^2|\Phi|^2 + \lambda|\Phi|^4\,.
\ee
%

It is convenient to split the field into the condensate (zero-momentum modes) and the excitations:
\be
\Phi = \frac{1}{\sqrt{2}}\big[(\rho_1 + \phi_1) + {\rm i}(\rho_2 + \phi_2)\big]\,,
\label{decomp 1}
\ee 
where~$\rho_1, \rho_2$ represent the condensate, and~$\phi_1, \phi_2$ are the excitations. Owing to the~$U(1)$ symmetry, we have freedom in choosing the condensate such that~$\rho_1^2 +\rho_2^2$ is a constant. For simplicity, let us fix
\be
\rho_1 = \rho_2 =  \rho\,. 
\ee
Expanding the potential in powers of the excitations, we have, at zeroth-order,
\be
U^{(0)} = m^2\rho^2 + \lambda\rho^4\,.
\ee
The first-order terms can be ignored as usual since they do not contribute to the effective action in the one-loop approximation. The second and the fourth-order terms are respectively given by
\begin{subequations}
\begin{align}
U^{(2)} &=  \frac{m^2 + 2\lambda\rho^2}{2}(\phi_1^2 + \phi_2^2) + \lambda\rho^2(\phi_1 + \phi_2)^2\,; \\
U^{(4)} &= \frac{\lambda}{4}\left(\phi_1^2 + \phi_2^2 \right)^2\,.
\end{align}
\end{subequations}
We will not be needing the third-order piece~$U^{(3)}$, since we will apply the HFB approximation.  

The HFB method is a self-consistent mean-field approximation scheme (see, {\it e.g.},~\cite{Bender:2003jk}). In this approach, the fourth-order terms are approximated by
\begin{subequations}
\begin{align}
\phi_1^4 &\simeq 6\phi_1^2\langle \phi_1^2 \rangle  - 3\langle \phi_1^2 \rangle^2\,;\\
\phi_2^4 & \simeq 6\phi_2^2\langle \phi_2^2  \rangle - 3\langle \phi_2^2 \rangle ^2\,;\\
\phi_1^2\phi_2^2 & \simeq \phi_1^2\langle\phi_2^2 \rangle  + \phi_2^2\langle \phi_1^2\rangle + 4\phi_1\phi_2\langle\phi_1\phi_2 \rangle - \langle\phi_1^2 \rangle\langle \phi_2^2 \rangle -2\langle \phi_1\phi_2 \rangle ^2\,,
\end{align}
\end{subequations}
where the expectation values account for both thermal and quantum corrections. These HFB-approximated terms can be added to the zeroth and second-order contributions to the potential, giving us
\begin{subequations}
\label{HFB corrected potential}
\begin{align}
U^{(0)}_{\rm HFB} &= m^2\rho^2 + \frac{\lambda}{4}\Big(4\rho^4 - 3\langle \phi_1^2 \rangle^2  - 3\langle \phi_2^2 \rangle ^2   -2 \langle\phi_1^2 \rangle\langle \phi_2^2 \rangle   - 4\langle \phi_1\phi_2 \rangle ^2 \Big)\,;  \\
\nonumber
U^{(2)}_{\rm HFB} &= \left(\frac{m^2}{2} + 2\lambda\rho^2 + \frac{3\lambda}{2}\langle\phi_1^2 \rangle + \frac{\lambda}{2}\langle\phi_2^2\rangle \right)\phi_1^2 + \left(\frac{m^2}{2} + 2\lambda\rho^2 + \frac{\lambda}{2}\langle\phi_1^2 \rangle + \frac{3\lambda}{2}\langle\phi_2^2\rangle \right)\phi_2^2\\
&+ \Big( \lambda\rho^2 + \langle\phi_1\phi_2 \rangle\Big)2\phi_1\phi_2\,.
\end{align}
\end{subequations}
Since this HFB potential is not $U(1)$ invariant, the symmetry of the original theory is lost. This means that the would-be Goldstone boson has become massive. This conundrum, which is of course an artifact of the field decomposition and HFB approximation, has been pointed out previously~\cite{Amelino-Camelia:1997xip, Petropoulos:1998gt, Lenaghan:1999si} and is one version of the so-called Hohenberg-Martin dilemma~\cite{HOHENBERG1965291}.

\section{Two chemical potentials}
\label{2 chemical potentials justification}

We have seen that the HFB approximation gives an effective action that fails to simultaneously satisfy the self-consistent mean-field equation while having a gapless Goldstone mode. As discussed in the Introduction, several solutions have been proposed in the literature to deal with this issue. In this paper, our strategy is to introduce a second chemical potential, generalizing the prescription of Yukalov {\it et al.} in a non-relativistic setting~\cite{YUKALOV2008461}. The approach is similar to the one used by~\cite{Pilaftsis:2013xna}, in which they modify the equations of motion to ensure both constraints are satisfied.

This framework entails that we have one chemical potential~$\mu_0$ for the condensate and a separate chemical potential~$\mu_1$ for the excitations. These encode the fact
that the fraction of particles in the condensate and the excitations are fixed at a given~$T$,~$V$ and~$N$. Mathematically, let us minimize the free energy~$F(N_0, N_1)$~\cite{Yukalov2011},
\be
\label{Free energy two chemical potentials}
\delta F = \frac{\partial F}{\partial N_0}\delta N_0 + \frac{\partial F}{\partial N_1}\delta N_1 = 0\,.
\ee
The derivative of the free energy with respect to the number of particles is the chemical potential:~$\mu_0 = \frac{\partial F}{\partial N_0}$ and~$\mu_1 = \frac{\partial F}{\partial N_1}$.
Furthermore, since the total number of particles is conserved,~$\delta N_0 = - \delta N_1$. Thus~\eqref{Free energy two chemical potentials} becomes
\be
(\mu_0 - \mu_1)\delta N_0 = 0\,.
\ee
If, in a phase transition at a particular critical temperature, the fraction of particles in either of the phases is not fixed,~$\delta N_0\neq 0$, then~$\mu_0 = \mu_1$. Instead, in our case the fraction of particles in the condensate and in excited states are each fixed,~$\delta N_0 = 0$, therefore the two chemical potentials are no longer required to be equal. In fact, we will see that they are indeed different for~$T < T_{\rm c}$ and become equal at~$T=T_{\rm c}$, when the condensate vanishes. 

This prescription of two chemical potentials will allow us to circumvent the Hohenberg-Martin dilemma discussed previously, with each chemical potential satisfying a different constraint. The condensate
chemical potential,~$\mu_0$, enforces the mean-field equation of motion, and the one associated with the excitations,~$\mu_1$, ensures that Goldstone's theorem is satisfied. 

To incorporate two chemical potentials, the generating functional~\eqref{Z one mu} is generalized to\footnote{Since $\rho_i$ correspond to the ground state, their Fourier transform gives the energy of the ground state which is zero. Hence, terms having $\dot{\rho_i}$ are ignored.  Terms having $\vec{\nabla}{\rho_i}$ are subsequently dropped for the same reason.}
\be
Z =\int \prod_{i=1}^2 \mathcal{D}\phi_i \mathcal{D}\pi_i \mathcal{D}\rho_i \mathcal{D}\chi_i \, \exp{\left[ \int_0^{\beta}{\rm d \tau}  \int {\rm d}^3x\left({\rm i}\pi_1\dot{\phi}_1 + {\rm i}\pi_2\dot{\phi}_2 - \mathcal{H} + \mu_0Q_0 + \mu_1Q_1 \right)\right]}\,,
\ee
where~$\pi_i$ and~$\chi_i$ are the momenta conjugate to~$\phi_i$ and~$\rho_i$, respectively. The conserved charge densities for the two phases are
\begin{subequations}
\begin{align}
Q_0 \equiv \rho_1\chi_2 - \rho_2\chi_1\,; \\
Q_1 \equiv \phi_1 \pi_2 - \phi_2\pi_1\,.
\end{align}
\end{subequations}
The Hamiltonian~$\mathcal{H}$ is given by
\be
\mathcal{H} = \frac{1}{2}\left(\pi_1^2 + \pi_2^2 + \chi_1^2 + \chi_2^2 + \big(\vec{\nabla}\phi_1\big)^2 +\big(\vec{\nabla}\phi_2\big)^2 \right) + U\big(|\Phi|^2\big)\,.
\ee
In the above expressions, we have used that~$\rho_i$ correspond to zero-momentum modes.

Performing the functional integrals over the conjugate momenta, fixing~$\rho_1 = \rho_2 = \rho$ as before, and applying the HFB approximation, we obtain
\be
    Z = \mathcal{N}\int\prod_{i=1}^2 \mathcal{D}\phi_i\mathcal{D}\rho \,\exp{\Big[-S[\phi_i,\rho] \Big]}\,,
\ee
where~$\mathcal{N}$ is an irrelevant multiplicative constant, and the action~$S$ is given by 
\begin{align}
\label{Action}
\nonumber
    S[\phi_i,\rho] = \int_0^{\beta} \text{d}\tau \int \text{d}^3x\ &\left[\frac{1}{2}\left(\dot{\phi}_1^2 + \dot{\phi}_2^2 + \big(\vec{\nabla}\phi_1\big)^2 + \big(\vec{\nabla}\phi_2\big)^2\right) + {\rm i}\mu_1\left(\dot{\phi}_1\phi_2- \phi_1\dot{\phi}_2\right)-\mu_0^2\rho^2  \right.\\
    &\left. +~U^{(0)}_{\rm HFB} + U^{(2)}_{\rm HFB}  -\frac{\mu_1^2}{2}\left(\phi_1^2 + \phi_2^2 \right) \right]\,,
\end{align}
where the HFB corrected potential terms,~$U^{(0)}_{\rm HFB}$ and~$U^{(2)}_{\rm HFB}$, are once again given by~\eqref{HFB corrected potential}. 
The Euler-Lagrange equation of motion for~$\rho$, given by~$\langle \frac{\delta S}{\delta \rho} \rangle = 0$, fixes the condensate chemical potential:
\be
\label{condensate chemical potential}
\mu_0^2 = m^2 + 2\lambda\Big(\rho^2 + \langle\phi_1^2\rangle + \langle\phi_2^2 \rangle+ \langle\phi_1\phi_2 \rangle\Big)\,.
\ee

Notice that the effective action~$S$ and the condensate chemical potential~$\mu_0$ both depend on the expectation values~$\langle\phi_1^2\rangle$,~$\langle\phi_2^2\rangle$ and~$\langle\phi_1\phi_2 \rangle$. 
These expectation values are set by the excitations, and therefore vanish at~$T=0$ (ignoring quantum effects). To proceed, therefore, we need to compute the one-loop corrections to the free energy,
which in turn will allow us to evaluate the relevant expectation values. Moving forward, we can eliminate $\rho$ from our calculation by using~\eqref{condensate chemical potential}. However, this substitution will make the equations more complicated. Hence, we will retain $\rho$ in our calculation while keeping in mind that it has already been fixed using the Euler-Lagrange equation of motion.

\section{One-loop free energy}
\label{Matsubara Summation}

In this Section we calculate the free energy density~\eqref{F def}, including finite temperature and quantum corrections at one loop.
The zeroth-order contribution can be read off from~\eqref{Action}:
\be
F^{(0)} =  -\mu_0^2\rho^2 + U^{(0)}_{\rm HFB} \,.
\label{F0}
\ee
To compute one-loop corrections, we focus on quadratic terms, written in the form
\be
S^{(2)} = \frac{1}{2}\int {\rm d}^4x{\rm d}^4y\, \phi_i(x) \mathcal{M}_{ij}\phi_j(y)\,.
\label{S2 Mij}
\ee
The matrix elements~$\mathcal{M}_{ij} = \frac{\delta^2 S}{\delta \phi_i(x)\phi_j(y)}$ are given by
\begin{subequations}
\label{Mass_Matrix}
\begin{align}
    \mathcal{M}_{11} =& \Big(-\partial^2 + m^2 + 4\lambda\rho^2 + 3\lambda\langle \phi_1^2 \rangle + \lambda\langle \phi_2^2\rangle - \mu_1^2  \Big)\delta^4(x-y)\,;\\
    \mathcal{M}_{12} =& \left(2{\rm i}\mu_1\frac{\partial}{\partial\tau} + 2\lambda\rho^2 + 2\lambda\langle\phi_1\phi_2 \rangle\right)\delta^4(x-y)\,;\\
    \mathcal{M}_{21} =& \left(-2{\rm i}\mu_1\frac{\partial}{\partial\tau} + 2\lambda\rho^2 + 2\lambda\langle\phi_1\phi_2 \rangle \right)\delta^4(x-y)\,;\\
    \mathcal{M}_{22} =& \Big(-\partial^2 + m^2 + 4\lambda\rho^2 + \lambda\langle \phi_1^2 \rangle + 3\lambda\langle \phi_2^2\rangle - \mu_1^2 \Big)\delta^4(x-y)\,.      
\end{align}
\end{subequations}
Our choice of setting~$\rho_1 = \rho_2$ in the field decomposition also leads us to expect that~$\langle\phi_1^2\rangle = \langle \phi_2^2\rangle$. As a consistency check, notice that 
the diagonal elements of~$\mathcal{M}$ become equal when~$\langle\phi_1^2\rangle = \langle \phi_2^2\rangle$, as expected. We will henceforth make this choice.

At this stage, we decompose the fields in Fourier modes. One option is to perform this decomposition in terms of the usual ladder operators. However, working with creation and annihilation operators becomes complicated at finite temperature and requires the formalism of thermofield dynamics~\cite{Matsumoto,Landsman:1986uw}. In this formalism, one introduces a dual field space, conjugate to the~$\phi_1$,~$\phi_2$ space. Furthermore, one must go beyond the imaginary time formalism by charting a path in the complex time plane, such that dynamical effects can be captured as evolution along the real-time axis, while thermal effects correspond to evolution along the imaginary axis. 

Since we are not interested in the dynamical nature of the relevant expectation values, we can avoid using thermofield dynamics altogether and instead Fourier decompose the fields as  
\be
\label{Fourier Decomp A}
\phi_i(\vec{x},\tau) = \sqrt{\beta}\sum_{n=-\infty}^{\infty}\int\frac{\text{d}^3k}{(2\pi)^3}\widetilde{\phi}_{i,n}\big(\vec{k}\big){\rm e}^{{\rm i}\left(\vec{k}\cdot\vec{x} + \omega_n\tau\right)},
\ee
with~$\omega_n = 2\pi n/\beta$. Substituting into~\eqref{S2 Mij}, the quadratic action becomes
\be
S^{(2)} = \frac{\beta^2}{2} \int\frac{\text{d}^3k}{(2\pi)^3}\sum_n \widetilde{\phi}_{i,-n}\big(-\vec{k}\big)\widetilde{\mathcal{M}}_{ij}\widetilde{\phi}_{j,n}\big(\vec{k}\big)\,,
\ee
where~$\widetilde{\mathcal{M}}$ is the momentum-space representation of~$\mathcal{M}$:
\begin{equation}
\widetilde{\mathcal{M}} = 
\begin{pmatrix}
\omega_n^2 +p_k  &  q + 2\mu_1\omega_n\\
q - 2\mu_1\omega_n& \omega_n^2 +p_k
\end{pmatrix}\,.
\label{mass matrix k space}
\end{equation}
Here we have defined
\bea
\nonumber
p_k  &\equiv&  k^2 +m^2 + 4\lambda\left(\rho^2 + \langle\phi_1^2 \rangle\right)  -\mu_1^2\,; \\
q &\equiv & 2\lambda \left(\rho^2 + \langle \phi_1\phi_2\rangle\right)\,.
\label{pk def}
\eea
It is convenient to transform to a new basis~$\widetilde{\psi}_{\pm}$  in which~$\widetilde{\mathcal{M}}$ is diagonal: 
\bea
\nonumber
\widetilde{\phi}_{1,n}\big(\vec{k}\big) & = &\ \frac{1}{\sqrt{2}}\left( \widetilde{\psi}_{+,n}\big(\vec{k}\big) - \sqrt{\frac{q+2\mu_1\omega_n}{q-2\mu_1\omega_n}}\widetilde{\psi}_{-,n}\big(\vec{k}\big) \right)\,;\\
\widetilde{\phi}_{2,n}\big(\vec{k}\big) &= &\ \frac{1}{\sqrt{2}}\left(  \sqrt{\frac{q-2\mu_1\omega_n}{q+2\mu_1\omega_n}} \widetilde{\psi}_{+,n}\big(\vec{k}\big) + \widetilde{\psi}_{-,n}\big(\vec{k}\big) \right) \,.
\label{phi psi 1}
\eea
The transformation is unitary, and its determinant is unity. In the diagonal basis, $S^{(2)}$ can be written as
\bea
\nonumber
S^{(2)} &=& \frac{\beta^2}{2}\int\frac{\text{d}^3k}{(2\pi)^3}\sum_n\left[\left(\omega_n^2+p_k + \sqrt{q^2 - 4\mu_1^2\omega_n^2}\right)\widetilde{\psi}_{+,n}\big(\vec{k}\big)\widetilde{\psi}_{+,-n}\big(-\vec{k}\big) \right.  \\
&& ~~~~~~~~~~~~~~~~~~~~\left.+ \left(\omega_n^2+p_k  - \sqrt{q^2- 4\mu_1^2\omega_n^2}\right)\widetilde{\psi}_{-,n}\big(\vec{k}\big)\widetilde{\psi}_{-,-n}\big(-\vec{k}\big) \right]\,.
\label{Action_diagonal_basis}
\eea
The coefficients, $\omega_n^2+p_k \pm \sqrt{q^2 - 4\mu_1^2\omega_n^2}$, are of course identified as the two eigenvalues of~$\widetilde{\mathcal{M}}$. Our goal in this Section is to compute the free energy, which only depends on the eigenvalues, and not on the exact form of the transformation. We will use the explicit transformation~\eqref{phi psi 1} in Sec.~\ref{Calculating the Expectation Values} to compute the expectation values~$\langle\phi_1^2 \rangle$ and~$\langle \phi_1\phi_2\rangle$. 

Letting~$Z^{(1-\text{loop})}$ denote the contribution of the one-loop corrections to the generating functional, we obtain
\bea
\nonumber
\ln Z^{(1-\text{loop})} &=& \frac{-V}{2}\sum_n\int\frac{\text{d}^3k}{(2\pi)^3}\left[ \ln\left(\frac{\omega_n^2+p_k + \sqrt{q^2 -4\mu_1^2\omega_n^2}}{T^2} \right) + \ln\left(  \frac{\omega_n^2+p_k  - \sqrt{q^2 -4\mu_1^2\omega_n^2}}{T^2}\right) \right] \\
\nonumber
&=& \frac{-V}{2}\sum_n\int\frac{\text{d}^3k}{(2\pi)^3} \ln \left( \frac{\left(\omega_n^2+p_k\right)^2 - q^2 + 4\mu_1^2\omega_n^2}{T^4}\right) \\
&=& \frac{-V}{2} \sum_n \int \frac{\text{d}^3k}{(2\pi)^3} \left[ \ln\left(\frac{\omega_n^2 + E_+^2(k)}{T^2} \right) +  \ln\left(\frac{\omega_n^2 + E_-^2(k)}{T^2} \right) \right]
\label{Generating Functional 1}
\eea
where the dispersion relations are given by
\be
E_{\pm}^2(k) = p_k +2\mu_1^2 \pm \sqrt{4\mu_1^2\left(p_k+\mu_1^2\right)  +q^2} \,.
\label{Dispersion Relations}
\ee
For each term in~\eqref{Generating Functional 1} we can perform the Matsubara summation using the identity
\be
\sum_n \ln\left(\frac{\omega_n^2 +E_{\pm}^2(k) }{T^2}\right) = \frac{E_{\pm}(k)}{T} + 2 \ln\left(1 - {\rm e}^{-\beta E_{\pm}(k)} \right) + \text{divergent constant}\,.
\ee  
Ignoring the divergent constant, we obtain the one-loop corrected free energy, 
\be
\boxed{F = F^{(0)} + T\int \frac{\text{d}^3k}{(2\pi)^3}\ln\left[\left( 1- {\rm e}^{- \beta E_+(k)}\right)\left(1 -{\rm e}^{ -\beta E_-(k)} \right)\right] + \int \frac{\text{d}^3k}{(2\pi)^3}\frac{E_+(k) + E_-(k)}{2}}\,.
\label{Free Energy}
\ee
The first term is the zeroth-order contribution~$F^{(0)}$ given in~\eqref{F0}, the second term is the 1-loop finite-temperature correction, and the last term is the zero-point energy. 
This form of the free energy~\eqref{Free Energy} is well known in prior literature, {\it e.g.},~\cite{Bernstein:1990kf}. The difference with our result is that the zeroth-order contribution
features a chemical potential that is different from the chemical potential entering the contribution from thermal excitations.



\section{Thermodynamic relations}
\label{Relativistic Equation of State}

In this Section we derive various thermodynamic relations. As a first step, we can fix the excitation chemical potential~$\mu_1$ by demanding
the existence of a gapless mode:~$\lim_{k \rightarrow 0}E_-(k) = 0$. Using~\eqref{Dispersion Relations}, we obtain 
\be
\mu_1^2 = m^2 + 2\lambda\Big( \rho^2 + \langle\phi_1^2\rangle +  \langle\phi_2^2\rangle -\langle\phi_1\phi_2 \rangle\Big)\,,
\label{excitation chemical potential}
\ee
where we have reintroduced~$\langle\phi_2^2\rangle$ by allowing for $\langle\phi_1^2\rangle \neq \langle\phi_2^2\rangle$. Comparing with the condensate chemical potential~$\mu_0$ in~\eqref{condensate chemical potential}, we see that the only difference between the two chemical potentials is due to~$\langle \phi_1\phi_2 \rangle$ being non-zero. However, setting this term to zero arbitrarily would sacrifice self-consistency. Had we started with just one chemical potential, the Euler-Lagrange equation for~$\rho$ would not be consistent with the existence of a gapless mode. This shows that the chemical potentials must in general be different in our approach. 


Using the usual thermodynamic relation~$n = -\partial F/\partial \mu$, we obtain the number density for the condensate and excitations,
\begin{subequations}
\bea
\label{Condensate_density_1}
n_0 & =&  -\frac{\partial F}{\partial \mu_0} =2\mu_0\rho^2\,;  \\
\label{Excitation_density_1}
n_1 &= & -\frac{\partial F}{\partial \mu_1} = - \sum_{e = \pm}\int\frac{\text{d}^3k}{(2\pi)^3}\left[\frac{\partial E_e}{\partial \mu_1}\left(\frac{1}{2} + \frac{1}{{\rm e}^{ \beta E_e } -1} \right) \right]\,,
\eea
\end{subequations}
with 
\be
\frac{\partial E_{e}}{\partial \mu_1} =  \frac{e\mu_1}{E_{e} \sqrt{4\mu_1^2\left(p_k+ \mu_1^2\right) + q^2}}\left(E_{e}^2  +p_k \right)\, ; \qquad (e = \pm)\,.
\ee
As temperature increases, particles will transition from the condensed phase to the excited phase, thereby reducing~$n_0$ while increasing~$n_1$, such that the total number density 
\be
\label{Number Density}
n = n_0 + n_1 
\ee
is conserved.

As a quick check on this result, consider the simplest example of an ideal Bose gas~($\lambda =0$). In this case, the dispersion relations~\eqref{Dispersion Relations} reduce to
\be
E_{\pm} = \sqrt{k^2 + m^2}\pm \mu_1\,.
\ee
Substituting into~\eqref{Excitation_density_1} gives
\be
n_1 = \int \frac{\text{d}^3k}{(2\pi)^3}\left(\frac{1}{{\rm e}^{ \beta E_-}-1} - \frac{1}{{\rm e}^{ \beta E_+}-1} \right)\,.
\label{Excitation Density bose fields}
\ee
We recognize the difference in the number of particles and anti-particles, which is the usual result for the net charge density in a relativistic field theory.
From this result, we can also see that the non-thermal part in~\eqref{Excitation_density_1} arises from contact interactions between particles. Since contact
interactions are an approximation to a potential that falls off with distance, we expect that our results will be UV-divergent. We will confirm later that the non-thermal part
indeed diverges and must be suitably regularized. 


\section{Expectation values and renormalization}
\label{Calculating the Expectation Values}

The expression for the free energy~\eqref{Free Energy}, as well as those for the chemical potentials and number densities, all depend on the expectation values~$\langle \phi_1^2 \rangle$ and~$\langle\phi_1\phi_2 \rangle$.
The goal of this Section is to calculate these expectation values. This will require the use of a renormalization scheme. Throughout the calculation we set~$\langle \phi_1^2\rangle = \langle\phi_2^2\rangle$, as justified below~\eqref{Mass_Matrix}.

\subsection{Correlation functions}

Our starting point is the transformation~\eqref{phi psi 1} from the original basis~$\widetilde{\phi}_{1,2}$ to the diagonal basis~$\widetilde{\psi}_{+,-}$. It implies 
\begin{subequations}
\bea
\label{phi1phi1n}
\widetilde{\phi}_{1,n}\widetilde{\phi}_{1,-n} &=& \widetilde{\phi}_{2,n}\widetilde{\phi}_{2,-n} = \frac{1}{2}\left( \widetilde{\psi}_{+,n}\widetilde{\psi}_{+,-n}  + \widetilde{\psi}_{-,n}\widetilde{\psi}_{-,-n}\right)\,;\\
\label{phi1phi2n}
\widetilde{\phi}_{1,n}\widetilde{\phi}_{2,-n}  &= & \frac{1}{2}\sqrt{\frac{q+2\mu_1\omega_n}{q-2\mu_1\omega_n}}\left( \widetilde{\psi}_{+,n}\widetilde{\psi}_{+,-n}  - \widetilde{\psi}_{-,n}\widetilde{\psi}_{-,-n} \right)\,,
\eea
\end{subequations}
where~$q$ was defined in~\eqref{pk def}, and we have suppressed the~$\vec{k}$ dependence in the fields with the understanding that~$\vec{k}$ has the same sign as~\textit{n}. Furthermore, note that we have dropped the cross terms~$\widetilde{\psi}_{+,n}\widetilde{\psi}_{-,-n}$, since we are interested in calculating expectation values, and~$\langle \psi_+\psi_- \rangle = 0$ in the diagonal basis.

To compute~$\langle \phi_1^2\rangle$ and~$\langle\phi_1\phi_2\rangle$, we first need to evaluate~$\langle \psi_+^2\rangle$ and~$\langle \psi_-^2\rangle$, the expectation values for the diagonal basis fields. These can be expressed as a Matsubara sum over the Fourier correlators~$\langle \widetilde{\psi}_{1,n} \widetilde{\psi}_{1,-n}  \rangle$ and~$\langle \widetilde{\psi}_{2,n} \widetilde{\psi}_{2,-n}  \rangle$, which in turn are easily determined by performing functional integrals with the generating functional using the action~\eqref{Action_diagonal_basis}. This gives
\bea
\label{diagonal basis n space}
\nonumber
\langle \psi_+^2\rangle &=& \frac{V}{T} \int\frac{\text{d}^3k}{(2\pi)^3} \sum_n \left\langle \widetilde{\psi}_{+,n}\big(\vec{k}\big) \widetilde{\psi}_{+,-n}\big(-\vec{k}\big)  \right\rangle =  \int\frac{\text{d}^3k}{(2\pi)^3} \sum_n\frac{T}{\omega_n^2 +p_k+ \sqrt{q^2 -4\mu_1^2\omega_n^2}}\,;\\
\langle \psi_-^2\rangle &=& \frac{V}{T} \int\frac{\text{d}^3k}{(2\pi)^3}\sum_n \left\langle\widetilde{\psi}_{-,n}\big(\vec{k}\big) \widetilde{\psi}_{-,-n}\big(-\vec{k}\big) \right\rangle = \int\frac{\text{d}^3k}{(2\pi)^3} \sum_n\frac{T}{\omega_n^2 +p_k - \sqrt{q^2 -4\mu_1^2\omega_n^2}}\,.
\eea

Let us first study~$\langle \phi_1^2\rangle$. From~\eqref{phi1phi1n} we see that 
\be
\label{phi2expectation}
\langle\phi_1^2 \rangle = \frac{1}{2} \Big(\langle\psi_+^2 \rangle + \langle \psi_-^2\rangle\Big) =  \int \frac{\text{d}^3 k}{\left( 2\pi\right)^3}\sum_n\frac{T\left( \omega_n^2 + p_k\right)}{\left( \omega_n^2 + p_k\right)^2 - q^2 + 4\mu_1\omega_n^2 }\,.
\ee
Notice that the square root present in the individual correlators~$\langle\psi_+^2 \rangle$ and~$\langle \psi_-^2\rangle$, disappears in their sum. This is important because a square root in the contour integral would give rise to two branch cuts, which would prevent us from using a contour that encompasses all relevant poles. Hence, even though computing the expectation values for $\psi_\pm^2$ is rather difficult, computing their sum is straightforward since all the singularities are poles.

\begin{figure}[!h]
\centering
\begin{tikzpicture}
\draw
      (-2.7,0) edge[-latex] node[at end, right]{$\operatorname{Re} \omega$} (2.7,0)
      (0,-2.7) edge[-latex] node[at end, right]{$\operatorname{Im} \omega$} (0,3.1)
    ;
\draw[red, thick, decoration = {markings, 
	mark = at position 0.2 with {\arrow{latex}}, 
	mark = at position 0.495 with {\arrow{latex}},
	mark = at position 0.7 with {\arrow{latex}},
	mark = at position 0.995 with {\arrow{latex}} 
}, postaction = {decorate} ] (-0.2,2.7) -- (-0.2, -2.7) -- (0.2, -2.7) -- (0.2, 2.7)-- (-0.2, 2.7);
\filldraw [blue] (0,0) circle (2pt);
\filldraw [blue] (0,0.5) circle (2pt);
\filldraw [blue] (0,1) circle (2pt);
\filldraw [blue] (0,1.5) circle (2pt);
\filldraw [blue] (0,2) circle (2pt);
\filldraw [blue] (0,2.5) circle (2pt);
\filldraw [blue] (0,-0.5) circle (2pt);
\filldraw [blue] (0,-1) circle (2pt);
\filldraw [blue] (0,-1.5) circle (2pt);
\filldraw [blue] (0,-2) circle (2pt);
\filldraw [blue] (0,-2.5) circle (2pt);
\node[right] at (0.2,-2.4) {\small$C$};
\end{tikzpicture}
\begin{tikzpicture}
\draw
      (-2.9,0) edge[-latex] node[at end, right]{$\operatorname{Re} \omega$} (3.0,0)
      (0,-2.7) edge[-latex] node[at end, right]{$\operatorname{Im} \omega$} (0,3.1)
    ;
\draw[red, thick, decoration = {markings, 
	mark = at position 0.3 with {\arrow{latex}}}, postaction = {decorate} ] (0.2,-2.7) -- (0.2, 2.7) ;
\draw[red, thick, decoration = {markings, 
	mark = at position 0.3 with {\arrow{latex}}}, postaction = {decorate} ] (-0.2,2.7) -- (-0.2, -2.7) ;
\draw[red, thick, decoration = {markings, 
	mark = at position 0.3 with {\arrow{latex}}}, postaction = {decorate}] (0.2, 2.7) arc (90 - atan(0.2/2.7):-90 + atan(0.2/2.7): 2.706);
\draw[red, thick, decoration = {markings, 
	mark = at position 0.3 with {\arrow{latex}}}, postaction = {decorate}] (-0.2, -2.7) arc (-90 - atan(0.2/2.7):-270 + atan(0.2/2.7): 2.706);
\filldraw [blue] (2,0) circle (2pt);
\filldraw [blue] (1,0) circle (2pt);
\filldraw [blue] (-2,0) circle (2pt);
\filldraw [blue] (-1,0) circle (2pt);
\node[below] at (1,0) {\small$E_-$};
\node[below] at (2,0) {\small$E_+$};
\node[below] at (-1,0) {\small$-E_-$};
\node[below] at (-2,0) {\small$-E_+$};
\node[above] at (2.5, 1.6) {\small$C_1$};
\node[below] at (-2.5, -1.6) {\small$C_2$};
\end{tikzpicture}
\caption{The contour~$C$ in the Left Panel encompasses all the poles of~$\coth \frac{\beta \omega}{2}$ and is used to write the Matsubara sum~\eqref{Poles 1} in terms of a contour integral. 
The blue dots indicate the poles of~$\coth \frac{\beta \omega}{2}$ at~$\omega = {\rm i}\omega_n = 2\pi {\rm i} n T$. Since the infinitesimal parts at the very the top and bottom of~$C$ are zero, we get the Right Panel by closing the contour with two semi-circular arcs~$C_1$ and~$C_2$ of infinite radius. These enclose the poles that correspond to~$\omega = \pm E_\pm$.}
\label{Contour Integrals}
\end{figure}
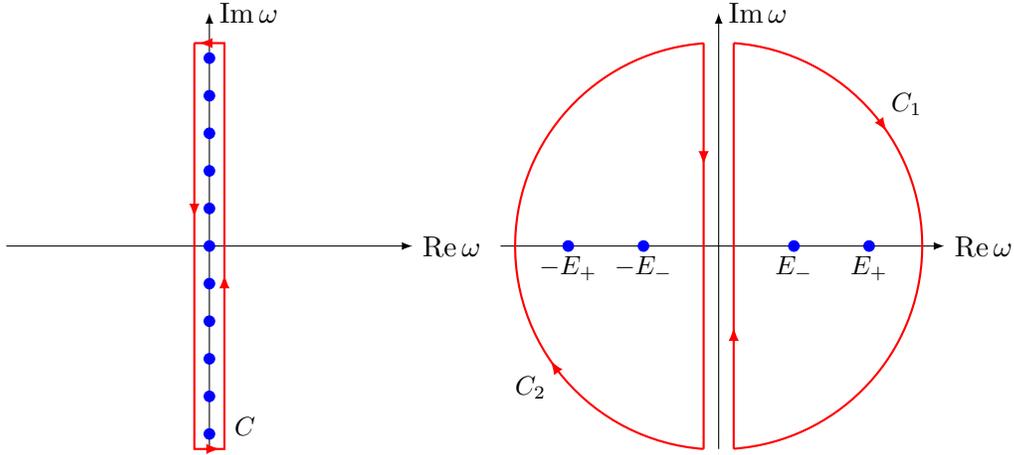

The Matsubara sum~\eqref{phi2expectation} can be performed through standard manipulations~\cite{Kapusta:2006pm}. We first write the sums as a contour integral encompassing the poles of~$\coth \frac{\beta \omega}{2}$:
\be
\label{Poles 1}
\sum_n\frac{T\left( \omega_n^2 + p_k\right)}{\left(\omega_n^2 + p_k\right) - \left(q^2 -4\mu_1^2\omega_n^2\right) } = \frac{1}{4\pi {\rm i}}\oint_C\text{d}\omega\frac{\left(-\omega^2 + p_k \right)}{\left(-\omega^2 + p_k\right)^2 - q^2 - 4\mu_1^2\omega^2 }\coth  \frac{\beta\omega}{2}\,.
\ee
As shown in the left panel of Fig.~\ref{Contour Integrals}, the contour~$C$ is oriented counter-clockwise and runs parallel on both sides of the imaginary~$\omega$-axis with infinitesimal segments crossing the imaginary axis and closing the contour at infinity on both sides. This contour encompasses all the poles of~$\coth \frac{\beta \omega}{2}$, which lie at
\be
\omega = {\rm i}\omega_n = 2\pi {\rm i} n T\,.
\ee
%
%

Since the two infinitesimal segments at the very top and bottom of~$C$ vanish, we can transform the contour~$C$ into semi-circular arcs~$C_1$ and~$C_2$ of infinite radius (right panel of Fig.~\ref{Contour Integrals}).
Contour~$C_1$ encompasses the poles at~$E_\pm$, while contour~$C_2$ encompasses~$-E_\pm$. The line integrals corresponding to the infinite semi-circular arcs are zero since the integrand scales as~$1/R^2$ as~$R \rightarrow \infty$, where~$R$ is the radius of the arc. Thus the Matsubara sum is equal to the integral over the contours~$C_1$ and~$C_2$ computed using the residue theorem. This gives
\be
\label{Phi1squared}
\langle\phi_1^2 \rangle  = \sum_{e=\pm}\int\frac{\text{d}^3k}{(2\pi)^3}\frac{1}{4E_e}  \left( 1 +  \frac{ 2e\mu_1^2  }{ \sqrt{4\mu_1^2\left(p_k+ \mu_1^2\right) + q^2}} \right)  \coth \frac{\beta E_e}{2} \,.
\ee

The expectation value~$\langle \phi_1\phi_2 \rangle$ is evaluated following similar steps. In this case we obtain
\be
\langle\phi_1\phi_2 \rangle = \int \frac{\text{d}^3 k}{\left( 2\pi\right)^3}\sum_n\frac{-T\left( 2\mu_1\omega_n + q\right)}{\left( \omega_n^2 + p_k\right)^2 - q^2 + 4\mu_1\omega_n^2}\,.
\ee
The part proportional to $\omega_n$ sums to zero because it is odd. The remaining sum is evaluated using the same methodology as before, with the result
\be
\label{Phi1Phi2}
\langle\phi_1\phi_2 \rangle = \sum_{e=\pm}\int\frac{\text{d}^3k}{(2\pi)^3}\frac{e}{2E_e}\frac{q}{\sqrt{4\mu_1^2\left(p_k+ \mu_1^2\right) + q^2}}    \coth \frac{\beta E_e}{2} \,.
\ee
The two equations~\eqref{Phi1squared} and~\eqref{Phi1Phi2} are implicit and coupled and must be solved together at a given temperature.
\subsection{Renormalization}
\label{Renormalization}

As alluded to earlier, the correlators~\eqref{Phi1squared} and~\eqref{Phi1Phi2}, as well as the zero-point energy term in~\eqref{Free Energy}, are all UV-divergent. We take care of these divergences using the renormalization scheme of~\cite{Alford:2013koa}. 

The divergence in~$\langle\phi_1^2 \rangle$ and~$\langle\phi_1\phi_2 \rangle$ arises from~$\coth(\beta E/2) = 1 + 2f_{\rm B}(E)$. The Bose factor approaches zero exponentially as~$k \rightarrow \infty$ and therefore gives a finite contribution, but the constant term is problematic. To cure this divergence, we introduce a hard momentum cut-off~$\Lambda$. We first separate out the temperature-dependent term, and, from the remaining expression, we then separate out the term with~$\mu_1 =0$. In other words, for the divergent part of the integrals for~$\langle\phi_1^2 \rangle$ and~$\langle\phi_1\phi_2 \rangle$, denoted respectively by~${\cal I}_{\phi_1^2}(\mu_1, \Lambda)$ and~${\cal I}_{\phi_1\phi_2}(\mu_1, \Lambda)$, we separate out the~$\Lambda$-dependent term as:
\be
{\cal I}(\mu_1,\Lambda) = {\cal I} (0,\Lambda) + {\cal I}(\mu_1)\,. 
\ee
The first part of these integrals is given by
\be
{\cal I}_{\phi_1^2}(0,\Lambda) = \sum_{e=\pm}\int \frac{\text{d}^3k}{(2\pi)^3}\frac{1}{4\nu_e}\,;\qquad  {\cal I}_{\phi_1\phi_2}(0,\Lambda) = \sum_{e=\pm}\int \frac{\text{d}^3k}{(2\pi)^3}\frac{e}{2\nu_e}\,,
\ee
where
\be
\nu_{\pm}^2 = p_k + \mu_1^2  \pm q\,.
\ee
Thus the renormalized correlators are
\bea
\nonumber
	\langle\phi_1^2 \rangle  &=& \sum_{e=\pm}\int\frac{\text{d}^3k}{(2\pi)^3} \left[ \frac{1}{4E_e}  \left( 1 +  \frac{ 2e\mu_1^2  }{\sqrt{4\mu_1^2\left(p+ \mu_1^2\right) + q^2}} \right)  \coth \left(\frac{\beta E_e}{2}\right)  -\frac{1}{4\nu_e}  \right]\,;   \\
	 \langle\phi_1\phi_2 \rangle &=&\sum_{e=\pm}\int\frac{\text{d}^3k}{(2\pi)^3} \frac{e}{2}  \left[ \frac{q }{E_e\sqrt{4\mu_1^2\left(p+ \mu_1^2\right) + q^2}}\coth \left(\frac{\beta E_e}{2}\right)  - \frac{1}{\nu_e} \right]\,.
\label{Renormalized Phis}
\eea

Looking at the free energy~\eqref{Free Energy}, we can similarly subtract from the zero-point energy the modified dispersion relation ($\nu_\pm$) obtained for~$\mu_1 =0$. The result is
%
\be
\label{Renormalized Free Energy}
\boxed{F = F^{(0)} +  T\int \frac{\text{d}^3k}{(2\pi)^3}\ln\left[\left( 1- {\rm e}^{-\beta E_+(k)}\right)\left(1 -{\rm e}^{-\beta E_-(k)} \right)\right] + \sum_{e=\pm}\ \int\frac{\text{d}^3k}{(2\pi)^3}\frac{E_e -\nu_e}{2}}\,.
\ee
The zeroth-order term~$F^{(0)}$ and finite-temperature terms are both finite, thanks to the renormalized correlators. For the zero-point energy term, the~$\nu_\pm$ subtraction removes the leading~$k$ divergence, but still leaves behind a divergent answer. One could add further counter-terms to make the result finite, as done in~\cite{Alford:2013koa}. Instead, to parallel the analysis in~\cite{Sharma:2018ydn}, we will deal with the zero-point integral in the non-relativistic regime using dimensional regularization in Sec.~\ref{Non-Relativistic Limit}. 

To summarize, for a given choice of~$m,\ \lambda,\ T$ and~$n$, we can now solve numerically the implicit expressions~\eqref{Renormalized Phis} for the renormalized correlators,~\eqref{condensate chemical potential} and~\eqref{excitation chemical potential} for the chemical potentials, as well as the requirement of charge conservation~\eqref{Number Density}. Substituting the solution to these equations in~\eqref{Renormalized Free Energy} gives the renormalized free energy.

\section{Non-Relativistic Limit}
\label{Non-Relativistic Limit}

As a check, we will work out the non-relativistic limit of our relativistic calculation to verify that the result is consistent with the non-relativistic analysis of~\cite{Sharma:2018ydn}. 

The non-relativistic chemical potentials~$\mu_{0,1}^{\rm NR}$ are related to their relativistic counterparts via~$\mu_{0,1}^{\rm NR} = \mu_{0,1} - m$. Using~\eqref{condensate chemical potential} and \eqref{excitation chemical potential} we obtain
\bea
\nonumber
\mu_0^{\rm NR} & \simeq & \frac{\lambda}{m} \Big(\rho^2 + \langle\phi_1^2\rangle + \langle\phi_2^2 \rangle+ \langle\phi_1\phi_2 \rangle\Big) \,;\\ 
\mu_1^{\rm NR} & \simeq & \frac{\lambda}{m} \Big(\rho^2 + \langle\phi_1^2\rangle + \langle\phi_2^2 \rangle - \langle\phi_1\phi_2 \rangle\Big) \,,
\label{NR chem pot}
\eea
where we have assumed that~$m^2  \gg \lambda \rho^2,\ \lambda\langle\phi_1^2\rangle$ and~$\lambda \langle\phi_1\phi_2\rangle$. Expanding the dispersion relations~\eqref{Dispersion Relations} for small~$k$ gives
\be
E_+(k) \simeq\  2m + \frac{k^2}{2m} \,;\qquad E_-(k) \simeq \ \sqrt{\frac{k^2}{2m} \left(\frac{k^2}{2m} + \frac{q}{m} \right) }\,.
\ee
Thus we recognize~$\psi_+$ as the massive mode, and~$\psi_-$ as the gapless mode. Furthermore, we can read off that the gapless mode has a linear dispersion relation for low momentum, with sound speed
\be
c_s^2 = \frac{q}{2m^2}\,.
\ee
This massless mode dominates the contribution to the excitations in the non-relativistic limit, so we henceforth ignore the contribution of the massive mode by setting~$\langle \psi_+^2\rangle \simeq 0$.  

The condensate number density~\eqref{Condensate_density_1} is approximately
\be
\label{non-rel condensate density}
n_0 \simeq 2m\rho^2\,.
\ee
Meanwhile, the excitation number density~\eqref{Excitation_density_1} is given by $n_1 = -\partial F/\partial \mu_1 = - \langle\delta S/\delta \mu_1\rangle$.
Ignoring the massive mode, each time derivative in the action~\eqref{Action} can be replaced by i$E_-$. Thus we obtain
\be
\label{non-rel excitation density}
n_1 \simeq m\Big(\langle \phi_1^2\rangle + \langle \phi_2^2 \rangle\Big)\,.
\ee
Therefore the non-relativistic chemical potentials~\eqref{NR chem pot} reduce to
\bea
\nonumber
\mu_0^{\rm NR} & \simeq &g\big(2n - n_0 + \sigma\big)\,; \\
\mu_1^{\rm NR} & \simeq & g \big(2n - n_0 - \sigma \big)\,,
\label{mu tom}
\eea
where~$g \equiv \frac{\lambda}{2m^2}$, and
\be
\sigma \equiv 2m\langle \phi_1\phi_2\rangle
\ee
is the so-called anomalous average. Equations~\eqref{mu tom} agree precisely with Eqs.~(42) and (44) in~\cite{Sharma:2018ydn}. 

\begin{figure}[htb!]
\centering   
\includegraphics[height=2.2in]{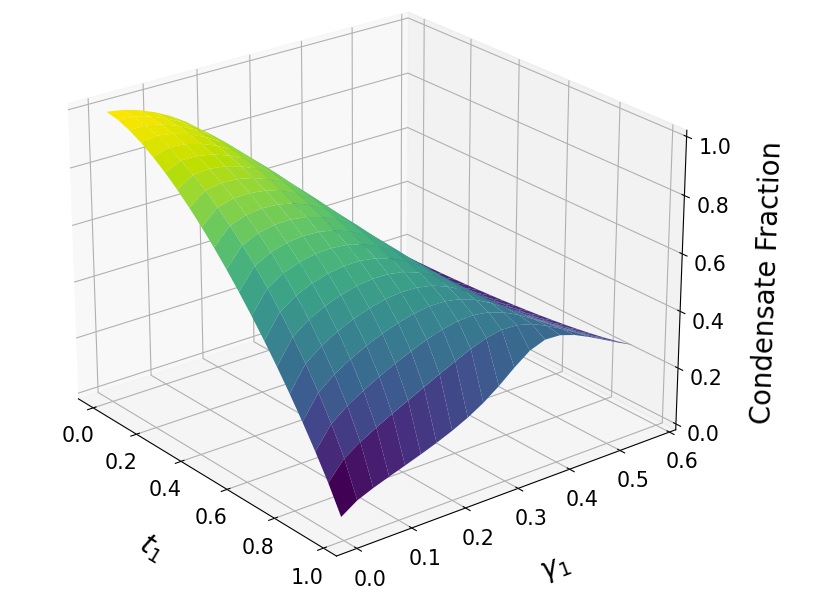}
\includegraphics[height=2.2in]{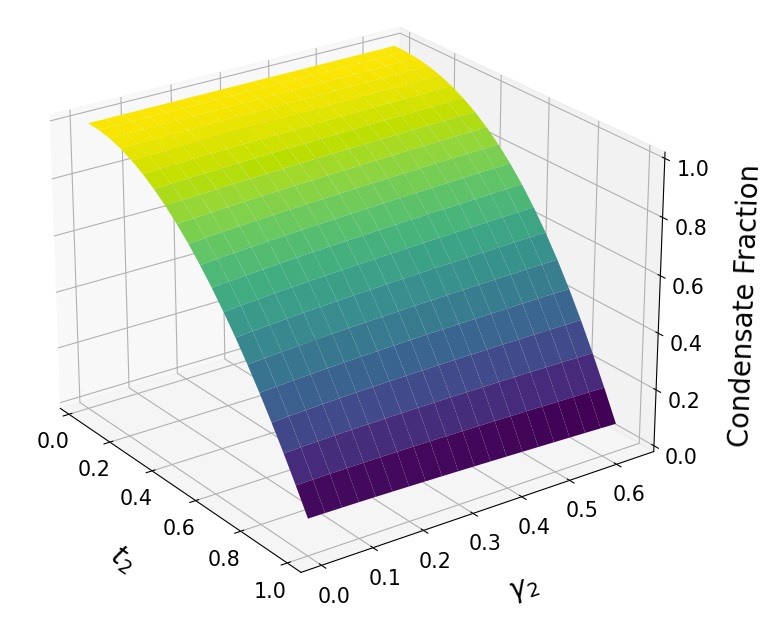}
\caption{Comparison of the condensate fraction~$\frac{n_0}{n} = 1 - \eta$ between the non-relativistic case with~$m = 0.5\,{\rm eV}$ and~$n = 10^{-9}~{\rm eV}^3$ (Left Panel), and the mildly relativistic case with~$m = 0.5\,{\rm eV}$ and~$n = 4\times 10^{-6}~{\rm eV}^3$ (Right Panel). The dimensionless temperatures~$t_1$ and~$t_2$ are normalized with respect to the critical temperature of an ideal Bose gas in the non-relativistic and mildly relativistic regimes, respectively~\cite{2007_Tc}. See the main text for details.} 
\label{3pt_Sf}
\end{figure}

Implicit expressions for~$n_1$ and~$\sigma$ can be obtained by substituting the renormalized correlators~\eqref{Renormalized Phis}. The latter also simplify in the
non-relativistic limit, with the result 
\bea
\label{Non_Rel_Excitation_Density_2}
n_1 &= &\int \frac{\text{d}^3k}{(2\pi)^3} \left[\frac{\frac{k^2}{2m} + g(n_0 + \sigma)}{2E_-}\coth \left(\frac{\beta E_-}{2} \right)  -\frac{1}{2}\right] \,; \\
\sigma &=& -\int \frac{\text{d}^3k}{(2\pi)^3} \frac{g(n_0 + \sigma)}{2E_-}\coth\left(\frac{\beta E_-}{2} \right)\,,
\eea
which is consistent with Eq.~(47) of~\cite{Sharma:2018ydn}. 

These implicit equations can be solved numerically, once we specify~$m,\ \lambda,\ T$ and~$n$. To do so, it is convenient to define dimensionless excitation and anomalous fractions:
\be
\eta = \frac{n_1}{n}\,;\qquad \xi = \frac{\sigma}{n}\,.
\label{eta xi}
\ee
The condensate fraction is just~$\frac{n_0}{n} = 1 - \eta$. Furthermore, instead of working with~$T$ and $\lambda$, we can define dimensionless temperature and interaction strength respectively as
\be
t_1 = \frac{mT}{3.31n^{2/3}\hbar^2}\,;\qquad \gamma_1 = \frac{\lambda n^{1/3}}{8\pi m}\,.
\ee
Note that~$t_1 = T/T_{\rm c}^{\rm NR}$, where~$T_{\rm c}^{\rm NR}$ is the critical temperature for a non-relativistic ideal Bose gas. The non-relativistic condensate fraction~$\frac{n_0}{n}$ is plotted in the left panel of Fig.~\ref{3pt_Sf} as a function of~$t_1$ and~$\gamma_1$, for~$m = 0.5\,{\rm eV}$ and~$n = 10^{-9}~{\rm eV}^3$. Over the range~$0 \leq t_1 \leq 1$ shown in the Figure,~$T/m$ ranges from~$0$ to~$10^{-6}$, which confirms the validity of the non-relativistic approximation.

For comparison, the right panel of Fig.~\ref{3pt_Sf} shows the condensate fraction in the mildly relativistic regime as a function of~$t_2 =  T  \sqrt{\frac{m}{3n}}$ and~$\gamma_2 = \frac{\lambda n^{2/3}}{m^2}$, with~$m = 0.5\,{\rm eV}$ and~$n = 4\times 10^{-6}~{\rm eV}^3$. Note that~$t_1 = T/T_{\rm c}^{\rm R}$, where~$T_{\rm c}^{\rm R}$ is the critical temperature for a relativistic ideal Bose gas. Over the range~$0 \leq t_2 \leq 1$ shown in the Figure,~$T/m$ ranges from~$0$ to $0.1$, corresponding to a mildly relativistic regime. Our 1-loop effective description breaks down for~$T \gtrsim mc_s^2$~\cite{Alford:2012vn}, such that we cannot reliably describe the ultra-relativistic regime. In the relativistic case, our framework breaks down for small~$m$ close to the critical temperature. Interestingly, we see from the Figure that increasing the interaction strength increases the condensate fraction significantly in the non-relativistic case, but has no noticeable effect on the condensate fraction in the mildly relativistic case.
 
The expression for the renormalized free energy~\eqref{Renormalized Free Energy} also simplifies in the non-relativistic limit.
In terms of the normal and the anomalous fractions~\eqref{eta xi}, the zeroth-order contribution~$ F^{(0)}$ becomes
\be
\label{E_0}
F^{(0)} = -\frac{\lambda}{2m^2}n^2\left(1+ \xi^2 - \frac{1}{2}(1-\eta-\xi)^2 \right)\,.
\ee
Meanwhile, the zero-point energy contribution is given by
\be
V_{\text{zero-point}} = \int \frac{\text{d}^3k}{(2\pi)^3}\left( \frac{E_+ - \nu_+}{2} +  \frac{E_- - \nu_-}{2}\right) \,.
\ee
As mentioned below~\eqref{Renormalized Free Energy}, this contribution remains divergent. To parallel the analysis in~\cite{Sharma:2018ydn}, we evaluate the momentum integral using  
dimensional regularization~\cite{Andersen:2003qj}. The dominant contribution in dimensional regularization comes from the gapless mode, which yields:
\be
\label{V_zero_point}
V_{\text{zero point}} = \frac{8m^{3/2}}{15\pi^2}\left(\frac{\lambda n}{2m^2}\right)^{5/2}\left(1-\eta+\xi\right)^{5/2}\,. 
\ee
Equations~\eqref{E_0} and~\eqref{V_zero_point} agree with Eqs.~(51) and (23) of~\cite{Sharma:2018ydn}, respectively. 

It is instructive to consider our results at~$T = 0$ and in the limit of a dilute Bose gas,~$a_s^3 n\ll 1$, where~$a _s= \frac{\lambda}{8\pi m}$ is the $s$-wave scattering length.
As argued in~\cite{Sharma:2018ydn}, in the limit~$T\rightarrow 0$ the normal density~$n_1$ goes to zero, but the anomalous average~$\sigma$ remains finite. In the
dilute limit the integral in~\eqref{Non_Rel_Excitation_Density_2} can be performed explicitly, with the result
\be
\xi = \frac{8}{\sqrt{\pi}} \sqrt{a^3n} + \ldots
\ee
Substituting into~\eqref{E_0} and~\eqref{V_zero_point}, we obtain the free energy at~$T = 0$  
\be
F(T=0) = -\frac{2\pi a_sn^2}{m}\left(1+ \frac{112}{15\sqrt{\pi}}  \sqrt{a_s^3 n} + \ldots\right)\,.
\ee
This differs from the result of Lee and Yang~\cite{Lee:1957zza},
\be
F_{\rm LY} (T=0) = -\frac{2\pi a_sn^2}{m}\left(1- \frac{128}{15\sqrt{\pi}}  \sqrt{a_s^3 n} + \dots\right)\,,
\ee
which ignores the contribution from the fourth-order terms. In our case, we have a non-zero anomalous average due to quantum corrections, which result from the fourth-order terms.

\section{Hydrodynamics of a Superfluid}
\label{Hydrodynamics of a Superfluid}

The existence of a BEC is related to the phenomenon of superfluidity, though there are some technical differences between the two~\cite{Khalatnikov:1965}. 
In Landau's phenomenological model, a superfluid at finite (sub-critical) temperature behaves as a mixture of two fluids~\cite{PhysRev.60.356}: an inviscid superfluid component, and a ``normal" component,
which is viscous and carries entropy. In this Section we will use the results above to split the field into superfluid and normal fluid components, and derive an explicit dictionary to the hydrodynamical description. 

For this purpose it is helpful to generalize the field decomposition~\eqref{decomp 1} to
\be
\Phi = \frac{1}{\sqrt{2}}\left(\rho {\rm e}^{{\rm i}\psi_0(x)} + (\phi_1 + {\rm i}\phi_2) {\rm e}^{{\rm i}\psi_1(x)} \right)\,.
\label{decomp 2}
\ee
Allowing the phases~$\psi_0$ and~$\psi_1$ to have spatial gradients enables the condensate and the excitations, respectively, to have finite velocity with respect to the frame of interest.\footnote{The parametrization~$(\phi_1 + {\rm i}\phi_2) {\rm e}^{{\rm i}\psi_1(x)}$ for the excitations is clearly redundant, but allows for a simple mapping to our earlier results by setting~$\psi_1 = -\mu_1 t$.} The gradient of the phases is proportional to the velocity of the superfluid in a particular frame, as we will see, and thus vanish in the rest frame of the superfluid. 

Instead of implementing the chemical potentials as Lagrange multipliers, it is convenient to include them as part of the phases. Concretely, the results of the previous Sections are recovered by setting~$\psi_\alpha(t) = -\mu_\alpha t$,~$\alpha=0,1$. The Lagrangian can once again be evaluated order by order in powers of the excitations. Ignoring odd-order terms, since they do not contribute in the HFB approximation, we only concern ourselves with even-order terms:
\begin{subequations}
\label{Lagrangian}
\begin{align}
\mathcal{L}^{(0)} = &- \rho^2\big(\partial_{\mu}\psi_0\partial^{\mu}\psi_0+ m^2\big)- \lambda\rho^4\,; \\
\mathcal{L}^{(2)} = & - \frac{1}{2}\big(\partial_{\mu}\phi_1 \partial^{\mu}\phi_1  +    \partial_{\mu}\phi_2 \partial^{\mu}\phi_2 \big) - \frac{\phi_1^2 + \phi_2^2}{2}\big(\partial_{\mu}\psi_1 \partial^{\mu}\psi_1  + m^2 + 4\lambda\rho^2\big)  \nonumber\\
& - \frac{1}{2}\partial_{\mu}\psi_1\big(\phi_1\partial^{\mu}\phi_2  - \phi_2\partial^{\mu}\phi_1  \big) +  2\lambda\rho^2\phi_1\phi_2\,;\\
\mathcal{L}^{(4)} = &-\lambda \big( \phi_1^2 + \phi_2^2\big)^2\,.
\end{align}
\end{subequations}
This reproduces the Hamiltonian of Sec.~\ref{section one-loop effective action} once we set~$\psi_\alpha(t) = -\mu_\alpha t$,~$\alpha=0,1$.

In the case of a single chemical potential, there is a single conserved current~\cite{Alford:2012vn}
\be
j^{\mu} = n_{\rm s}\frac{\partial^{\mu}\psi }{\chi} +  n_{\rm n} u^{\mu}\,,
\ee
with~$\chi = \sqrt{-\partial_{\mu}\psi\partial^{\mu}\psi}$, and where~$n_{\rm s}$ and~$n_{\rm n}$ are the number density for the superfluid and normal components, respectively.
Meanwhile,~$u^\mu$ is the four-velocity of the normal component. This current satisfies the usual continuity equation~$\partial_{\mu}j^{\mu} = 0$. 

In our approach with two chemical potentials, there are two conserved currents, given by~\cite{Son:2002zn}
\be
j_\alpha^{\mu} = n_{{\rm s}_\alpha}\frac{\partial^{\mu}\psi_\alpha }{\chi_\alpha} + n_{{\rm n}_\alpha} u^{\mu}\,;  \qquad \alpha = 0,1\,,
\ee
with~$\chi_\alpha = \sqrt{-\partial_{\mu}\psi_\alpha\partial^{\mu}\psi_\alpha}$. These reflect our demand that the charge in the  
condensate and excited states are individually conserved at fixed temperature. The total superfluid and normal component densities are given by
\be
n_{\rm s} = n_{{\rm s}_0} + n_{{\rm s}_1}\,;\qquad n_{\rm n} = n_{{\rm n}_0} + n_{{\rm n}_1} \,.
\ee
Thus, in general,~$n_{\rm s}$ and~$n_{\rm n}$ receive contributions from both the condensate and the excitations. 

In the normal fluid rest frame, where~$u^{\mu} = (1,0,0,0)$, the currents become~$\vec{\jmath}_\alpha = n_{{\rm s}_\alpha} \frac{\vec{\nabla} \psi_\alpha}{\chi_\alpha}$,
which implies
\be
n_{{\rm s}_\alpha} = \chi_\alpha   \frac{\vec{\nabla} \psi_\alpha \cdot \vec{\jmath}_\alpha}{\big(\vec{\nabla} \psi_\alpha\big)^2} \,.
\label{ns alpha}
\ee
On the other hand, the conserved currents derive from the free energy via Noether's theorem,
\be
j_\alpha^{\mu} = \frac{\partial F}{\partial (\partial_\mu\psi_\alpha)}\,.
\label{j from F}
\ee
In the limit of small superflow, which is the regime of interest, we can expand the free energy as
\be
F = F_0 + \frac{1}{2} \big(\vec{\nabla}\psi_\alpha \big)^2 \left.\left(\frac{\partial^2 F}{\partial \big\vert\vec{\nabla}\psi_\alpha\big\vert^{\,2}} \right)\right\vert_{\vec{\nabla}\psi_\alpha =0}\,,
\ee
where we have used the fact that~$F$ only depends on~$\big(\vec{\nabla}\psi_\alpha \big)^2$. Thus~\eqref{j from F} gives
\be
\vec{\jmath}_\alpha \simeq \vec{\nabla}\psi_\alpha \left.\left(\frac{\partial^2 F}{\partial \big\vert\vec{\nabla}\psi_\alpha\big\vert^{\,2}} \right)\right\vert_{\vec{\nabla}\psi_\alpha =0}\,.
\ee
Substituting this into~\eqref{ns alpha}, we obtain
\be
\label{Superfluid_density_1}
\boxed{n_{{\rm s}_\alpha} \simeq \mu_\alpha \left.\left(\frac{\partial^2 F}{\partial \big\vert\vec{\nabla}\psi_\alpha\big\vert^{\,2}} \right)\right\vert_{\vec{\nabla}\psi_\alpha =0}}\,,
\ee
where we have used~$\chi_\alpha\simeq \mu_\alpha$ at this order. This expression differs from the result obtained with one chemical potential~\cite{Alford:2012vn}, but agrees with it once we set~$\mu_0 = \mu_1$ and~$\psi_0 = \psi_1$.

To compute the superfluid densities~$n_{{\rm s}_\alpha}$ explicitly, we must generalize the free energy to include spatial gradients of the phases. The dependence on~$\big(\vec{\nabla}\psi_0\big)^2$ comes solely from the zeroth-order term~$F^{(0)}$. It is easy to see that~\eqref{F0} generalizes to 
\be
F^{(0)} = - \left(\mu_0^2 - \big(\vec{\nabla}\psi_0\big)^2\right) \rho^2 + U^{(0)}_{\rm HFB} \,,
\ee
which implies
\be
n_{{\rm s}_0} \simeq 2\mu_0\rho^2\,.
\label{ns0}
\ee
This is recognized as the condensate density~$n_0$ obtained in~\eqref{Condensate_density_1}, which tells us that the condensate only contributes to the superfluid component ({\it i.e.},~$n_{{\rm n}_0} = 0$).

Meanwhile, the dependence on~$\big(\vec{\nabla}\psi_1\big)^2$ comes from the 1-loop corrections. These take the same form as in~\eqref{Free Energy}, but with modified~$E_\pm (k)$ to account for non-zero superflow. Specifically, we obtain
\be
n_{{\rm s}_1} \simeq  \mu_1\sum_{e=\pm}\int \frac{\text{d}^3k}{(2\pi)^3}\left[  \frac{\partial^2E_e}{\partial|\vec{\nabla}\psi_1|^2} \left(\frac{1}{2} + f_{\rm B}(E_e) \right) 
-  \left(\frac{\partial E_e}{\partial |\vec{\nabla}\psi_1|}\right)^2     \beta f_{\rm B}(E_e)\Big(1+f_{\rm B}(E_e)\Big)   \right]_{\vec{\nabla}\psi_1 =0} \,.
\label{ns1}
\ee
Thus it remains to calculate~$\frac{\partial E_\pm(k)}{\partial |\vec{\nabla}\psi_1|}\Big\rvert_{\vec{\nabla}\psi_1 = 0}$ and~~$\frac{\partial^2 E_\pm(k)}{\partial |\vec{\nabla}\psi_1|^2}\Big\rvert_{\vec{\nabla}\psi_1 = 0}$. 
To do so, we go back to the mass matrix~$\mathcal{M}$. Allowing for spatial gradients of the phases,~\eqref{mass matrix k space} generalizes to 
\begin{equation}
\widetilde{\mathcal{M}} = 
\begin{pmatrix}
\omega_n^2+ p_k  + \chi_1^2  &  2{\rm i} k^\mu \partial_\mu\psi_1  + q\\
- 2{\rm i} k^\mu \partial_\mu \psi_1  + q &\omega_n^2+p_k  +  \chi_1^2 
\end{pmatrix}\,,
\end{equation}
with~$k^\mu \equiv ({\rm i}\omega_n,\vec{k})$, and where~$p_k$ and~$q$ are defined in~\eqref{pk def}.
Similar to the calculation of the previous Section, the vanishing of the determinant gives the dispersion relations: 
\be
\label{Modified_dispersion}
\big(-\omega^2+ p_k  + \chi_1^2 \big)^2 - q^2 - 4(k^\mu\partial_\mu\psi_1)^2 = 0\,,
\ee
where we have used~$\omega = {\rm i}\omega_n$. The solution gives the desired dispersion relations:
\be
E_{\pm}^2(k) = p_k +\chi_1^2 \pm \sqrt{4(k^\mu\partial_\mu\psi_1)^2  +q^2}   \,.
\ee
This is an implicit relation, however, because~$k^0 = E_{\pm}(k)$ in the above. Thus a closed form solution is difficult to obtain. Fortunately, the relevant quantities, {\it i.e}, the derivatives
of~$E_\pm$ with respect to~$ |\vec{\nabla}\psi_1|$, are easy to extract:
\bea
\nonumber
\frac{\partial E_\pm(k)}{\partial |\vec{\nabla}\psi_1|}\Bigg\rvert_{\vec{\nabla}\psi_1 = 0} & = &\pm \frac{2\mu_1k_{\parallel}}{A_k}\,;\\
\frac{\partial^2 E_\pm (k)}{\partial |\vec{\nabla} \psi_1|^{\,2}}\Bigg\rvert_{\vec{\nabla}\psi_1 = 0 } & = &  \frac{1}{E_\pm  A_k}  \left[ \pm \left(E_\pm^2 - 2k_{\parallel}^2 -k^2\right) +  \frac{8\mu_1^2k^2}{A_k} \mp  \frac{4\mu_1^2k^2\left(2E_\pm^2 \pm A_k \right)}{A^2_k}\right]\,,
\label{phase_derivatives}
\eea
where~$A_k \equiv  \sqrt{4\mu_1^2\left(p_k+ \mu_1^2\right) + q^2}$, and~$k_\parallel = \frac{\vec{k}\cdot \vec{\nabla}\psi_1}{\big\vert \vec{\nabla}\psi_1\big\vert}$. All quantities on the right-hand side are evaluated at vanishing spatial gradients, {\it e.g.}, with~$E_\pm(k)$ given by~\eqref{Dispersion Relations}. Substituting into~\eqref{ns1} gives the excitation contribution to the superfluid density.

The total superfluid density, to leading order in the superflow, is given by the sum of~\eqref{ns0} and~\eqref{ns1}:
\be
\boxed{n_{\rm s}  = \ 2\mu_0\rho^2 + \mu_1\sum_{e=\pm}\int \frac{\text{d}^3k}{(2\pi)^3}\left[ \frac{\partial^2E_e}{\partial|\vec{\nabla}\psi_1|^2}\left(\frac{1}{2} + f_{\rm B}(E_e) \right) - \left(\frac{\partial E_e}{\partial |\vec{\nabla}\psi_1|}\right)^2 \beta f_{\rm B}(E_e)\big(1+f_{\rm B}(E_e)\big)   \right]_{\vec{\nabla}\psi_1 =0}} \,.
\label{ns tot}
\ee
This expression greatly simplifies in the non-relativistic regime, where the massive excitations can be neglected and hence~$f_{\rm B}(E_+) \simeq 0$. Furthermore, in this regime it is easy to show from~\eqref{phase_derivatives} that~$\frac{\partial^2 E_+ (k)}{\partial |\vec{\nabla} \psi_1|^{\,2}}\Big\rvert_{\vec{\nabla}\psi_1 = 0}$ gives a suppressed contribution. Thus~\eqref{ns tot} becomes
\bea
\label{Superfluid_Non-Rel}
\nonumber
n_{\rm s} &\simeq& 2\mu_0\rho^2 + \frac{\mu_1}{2} \int \frac{\text{d}^3k}{(2\pi)^3} \frac{\partial^2E_-}{\partial|\vec{\nabla}\psi_1|^2}\Bigg\rvert_{\vec{\nabla}\psi_1 = 0}  \\
\nonumber
&+&  \mu_1\int \frac{\text{d}^3k}{(2\pi)^3} \frac{\partial^2E_-}{\partial|\vec{\nabla}\psi_1|^2}\Bigg\rvert_{\vec{\nabla}\psi_1 = 0}  f_{\rm B}(E_-) 
- \mu_1\beta\int \frac{\text{d}^3k}{(2\pi)^3} \left( \frac{\partial E_-}{\partial |\vec{\nabla}\psi_1|} \right)^2 f_{\rm B}(E_-)\big(1+f_{\rm B}(E_-))\,.\\
\eea
The different terms are to be interpreted as follows:

\begin{itemize}

\item As already mentioned, the first term is recognized as the condensate density~\eqref{Condensate_density_1},~$n_0 = 2\mu_0\rho^2$. 

\item The second term (on the first line) is independent of~$T$ and represents the contribution due to quantum corrections in the form of contact interactions at~$T=0$. It is easy to show that it matches the~$T=0$ part of the excitation density~$n_1$ given by~\eqref{Non_Rel_Excitation_Density_2}.\footnote{This can also be seen by substituting~\eqref{Superfluid_density_1} directly in~\eqref{Lagrangian} instead of first computing the effective action. This tells us that 
\be
\nonumber
-\mu_1 \left(\frac{\partial^2 \mathcal{L}}{\partial |\vec{\nabla}\psi_1|^2} \right)\Bigg\rvert_{\vec{\nabla}\psi_1 =0} = 2\mu_0\rho^2 + \mu_1\big(\langle\phi_1^2\rangle +\langle \phi_2^2 \rangle\big) + \text{terms that vanish as}~T\rightarrow 0\,.
\ee
}

\item Lastly the second line in~\eqref{Superfluid_Non-Rel} vanishes exponentially as~$T\rightarrow 0$ thanks to the Bose factors. 

\end{itemize}

It follows that, in the limit~$T\rightarrow 0$, the superfluid density is equal to the sum of the condensate density plus the excitation density:
\be
n_{\rm s} = n_0 + n_1 = n \ \ \text{at }T=0\,,
\ee
where we have used~\eqref{Number Density}. Thus the superfluid fraction is equal to unity at zero temperature, and there are no particles in the normal phase. This also agrees with the experimental observation that liquid helium, which can be modeled as having strong interactions, has a superfluid fraction close to 1, while the condensate fraction is~${\cal O}(10\%)$~\cite{PhysRevLett.21.787} since the excitation density increases with interaction strength. Thus, while the condensate depletes as the interaction strength between particles increases, there is no corresponding depletion of the superfluid. 

Along these lines, it also follows from~\eqref{Superfluid_Non-Rel} that the only way for the superfluid to deplete is through the temperature-dependent terms in the second line. These terms grow with increasing temperature, resulting in a depletion of the superfluid. In particular, for sufficiently large temperature (or sufficiently weak interaction strength), the second term in~\eqref{Superfluid_Non-Rel} arising from quantum corrections can be neglected compared to the thermal corrections. In this case, we obtain
\be
n_{\rm s} = 2\mu_0\rho^2 -  \frac{\beta}{12m}\int \frac{\text{d}^3k}{(2\pi)^3}\frac{k^2}{\sinh^2\left(\frac{\beta E_-(k)}{2} \right)}\,.
\ee
This matches the known result in the non-relativistic limit and for weak coupling, as shown in Eq.~(66) of~\cite{Sharma:2018ydn}.

\section{Outlook}

The problem of describing a BEC through a scalar field exhibiting spontaneous symmetry breaking has been well-known for over five decades in the condensed matter community.  After the development of the CJT formalism for studying self-consistent QFTs, this problem was again noted in terms of the inability to simultaneously satisfy the Euler-Lagrange equation of motion and Goldstone's theorem. A number of different approaches to solve this problem have been proposed over the years. Each offers different insights into the problem, but also usually suffers from a pathology or carries some undesirable baggage in the form of additional ad hoc terms or constraints.

Our own motivation for revisiting this problem is the recent interest in BEC and superfluid candidates for dark matter. This paper is a natural follow-up to our earlier work~\cite{Sharma:2018ydn}, where 
we derived the non-relativistic, finite-temperature equation of state for dark matter superfluids, using a self-consistent mean-field approximation. In this paper we extended the calculation using a relativistic 
QFT framework. As in~\cite{Sharma:2018ydn}, we followed Yukalov's proposal of using two chemical potentials to describe a BEC ---  one for the condensed phase, 
a second one for the normal phase. The two chemical potentials allow us to simultaneously satisfy the self-consistency condition for the mean-field while having 
a gapless Goldstone mode. 

Our main results can be summarized as follows. We applied this proposal in the context of an imaginary time formalism QFT to describe thermal effects. We worked out the free energy of the system, incorporating the renormalization scheme of~\cite{Alford:2013koa}.  Since our calculation was done self-consistently, the resulting expressions for the condensate (excitation) and anomalous densities are implicit and can be evaluated numerically. We then worked out the non-relativistic limit and showed its consistency with the earlier results in~\cite{Sharma:2018ydn}. Finally, we sought to clarify the relationship between superfluidity and BEC by translating our results to the hydrodynamical language and working out the superfluid fraction.

Though we performed an explicit calculation for a~$|\Phi|^4$ theory, our analysis can be easily generalized to any theory with a potential having~$|\Phi|^{2n}$ terms. It would be illuminating to repeat the analysis for the more realistic superfluid effective theory proposed in~\cite{Berezhiani:2015bqa, Berezhiani:2015pia,Berezhiani:2017tth}, with hexic potential. It would be interesting, in particular, to study various observable consequences of dark matter superfluidity in our language, such as the effect of core fragmentation~\cite{Berezhiani:2021rjs}.

Even though our solution is naturally framed in a way that makes it easier to map it to the physics of a BEC, it can be easily checked against other results in the literature. Comparing with the results of~\cite{Alford:2013koa}, for instance, we find that our calculation yields the same results as the usual CJT calculation, with the only differences arising from our choice of the two chemical potentials. This choice allows us to avoid the ad hoc method used in that particular calculation, as well as others, by introducing a physically well-motivated scheme of two chemical potentials. A similar method is also used in~\cite{Pilaftsis:2013xna}, wherein a Lagrange multiplier is introduced to define a new, truncated 2PI effective action, which essentially serves the same purpose as our second chemical potential. Understanding the origin of the various approaches to this problem, as well as their similarities/differences, can help us provide deeper insights into its resolution.

\vspace{.4cm}
\noindent
{\bf Acknowledgements:} We thank Lasha Berezhiani for initial collaboration on this project and for many helpful discussions. We also thank Shantanu Agarwal for providing us with interesting insights regarding contour integration in the presence of branch cuts.
This work is supported by the US Department of Energy (HEP) Award DE-SC0013528 and NASA ATP grant 80NSSC18K0694.

\renewcommand{\em}{}
\bibliographystyle{utphys}
\addcontentsline{toc}{section}{References}
\bibliography{ms.bib}

\end{document}